\documentstyle[11pt,newpasp,twoside,epsf]{article}
\def\edcomment#1{\iffalse\marginpar{\raggedright\sl#1\/}\else\relax\fi}
\reversemarginpar

\begin{document}
\title{Oxygen in the Galactic thin and thick disks}
 \author{T. Bensby, S. Feltzing, and I. Lundstr\"om}
\affil{Lund Observatory, Box 43, SE-22100 Lund, Sweden}

\begin{abstract}
First results from a study into the abundance trends of oxygen in the Galactic thin and thick
disks are presented. Oxygen abundances for 21 thick disk and 42 thin disk F and G dwarf stars 
based on very high resolution spectra ($R\sim215\,000$) and high signal-to-noise ($S/N>400$) 
of the faint forbidden oxygen line at 6300 {\AA} have been determined. We find that $\rm [O/Fe]$ 
for the thick disk stars show a turn-down, i.e. the ``knee'', at [Fe/H] between $-0.4$ and 
$-0.3$ dex indicating the onset of SNe type Ia. The thin disk stars on the other hand show a 
shallow decrease going from $\rm [Fe/H] \sim -0.7$ to the highest metallicities with no 
apparent ``knee'' present indicating a slower star formation history.
\end{abstract}

%===========================================================================================
\section{Introduction}

The Galactic thin and thick disks are two distinct stellar populations in terms of 
age distributions and kinematics. The chemical trends in the two systems are also most likely
different although recent works give conflicting results, see e.g. Chen et al.~(2000) and
Fuhrmann~(1998). We show that the abundance trends for oxygen are different
for the thin and thick disks.  

%==========================================================================================
\section{Observations}

The selection of thin and thick disk stars was based on kinematics and is fully described in 
Bensby et al.~(2003a, in prep). We calculated Gaussian 
probabilities for each star that it belongs to the thin and thick disk 
respectively, using the galactic velocity components $U$, $V$, and $W$ of the stars. Stars 
with high probabilities of belonging to either the thin or the thick disk were then 
selected. The sample consists of 21 thick disk stars and 42 thin disk stars.

Spectra were obtained with the CES spectrograph on the ESO 3.6m telescope with a
a resolution of $R\sim 215\,000$ and a signal-to-noise $S/N > 400$. Telluric lines were 
divided out using spectra from fast rotating B stars. Further details are given in Bensby et 
al.~(2003b, in prep).

%==========================================================================================
\section{Abundances and results}

Oxygen abundances were determined through fitting of synthetic spectra to the observed
spectra. The forbidden oxygen line at 6300 {\AA} that has a blend of nickel
in its right wing. At low metallicities this blend is often negligible, but becomes severe
at higher metallicities. This is illustrated in Fig.~1 where we plot synthetic and observed
spectra for three stars at different metallicities. Fe and Ni abundances have been determined 
from our FEROS spectra (R$\sim 48\,000$) by measuring equivalent widths of approximately 140 Fe\,{\sc i}, 
30 Fe\,{\sc ii}, and 50 Ni\,{\sc i} lines for each star (Bensby et al.~2003a in prep.) 

\begin{figure}[ht]
\plotone{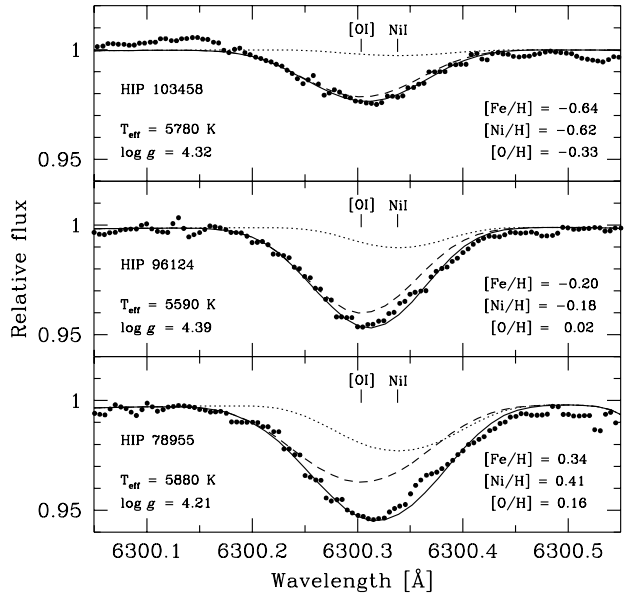}
\caption{The forbidden oxygen line at 6300 {\AA} for three stars with different metallicities;
         HIP 103458 (thick disk), HIP 96124 (thick disk), and HIP 78955 (thin disk). The observed
         spectra are plotted with solid circles. Three different synthetic spectra are 
         shown for each star:
         only the forbidden oxygen line (dashed line), only the blending nickel line (dotted line), 
         combination of the two (solid line).}
\end{figure}
\begin{figure}[ht]
\plotone{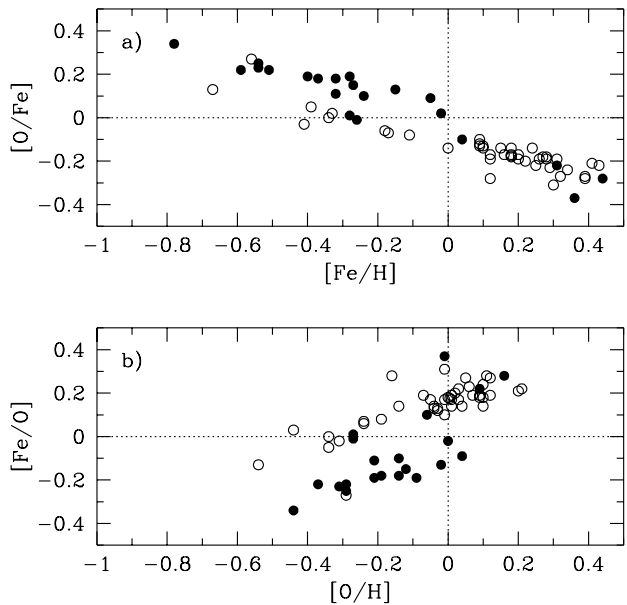}
\caption{Abundance trends for oxygen. Thick disk stars are marked by filled circles and
         thin disk stars by empty circles.}
\end{figure}

The two plots in Fig.~2 presents our results. These are our findings: \\
{\bf 1.} The thin and thick disk stars clearly show different abundance trends. This
      is a strong indication of their disparate origin and different epochs of formation. \\
{\bf 2.} A turn-down for $\rm [O/Fe]$ at $\rm [Fe/H]\sim -0.35$ for the thick disk stars,
      from being roughly flat, continuing down to solar values. This feature is most likely
      a signature of the onset of SNIa. \\
{\bf 3.} The thin disk stars show a shallow decrease when going from the lowest metallicities
      to solar values, {\it not} showing a knee. This implies that the star formation rate in 
      the thin disk was quite low compared to that in the thick disk. \\
{\bf 4.} At super-solar metallicities the trend found at sub-solar metallicities continues 
      linearly for the thin disk stars.
      In contrast Nissen and Edvardsson~(1992) found $\rm [O/Fe]$ to level out at
      these metallicities. However, they did not take the Ni\,{\sc i} blend in the [O\,{\sc i}] line
      into account, which becomes important at these metallicities, see Fig.~1.
      This result has implications for different models of supernova yields, and will be
      investigated further.

All stars have also been observed with the FEROS spectrograph and abundances for other
elements have been determined (Na, Mg, Al, Si, Ca, Sc, Ti, V, Cr, Mn, Fe, Co, Ni, Zn, Y,
Ba, Eu). For the $\alpha$-elements we find the same signature from the onset of SNIa in the
thick disk which appears to be absent in the thin disk, see Feltzing et al.~(2002) and
Bensby et al.~(2003a, in prep), in good agreement with the trends we find for oxygen.

A few stars merits, due to their positions in Fig.~2, further comments: two thick disk stars at
$\rm [Fe/H]\sim-0.3$ and one thin disk star at $\rm [Fe/H]\sim-0.6$. The latter may
be due to the fact that the thick disk also contain stars with ``cold'' kinematics. The first
two are a bit harder to understand but their kinematics might have been heated through
close encounters or they might have been kicked-out from a double or multiple stellar
system.

%=========================================================================================

%==========================================================================================

\end{document}